\documentclass[oneside,reqno,a4paper,12pt]{amsart}

\usepackage{latexsym}
\usepackage{euscript}
\usepackage{epsfig}

\newcommand{\be}{\begin{equation}}
\newcommand{\ee}{\end{equation}}

\def\ie{{\it i.e.}}
\def\LEP2{{LEPII}}

\begin{document}

\title[]{CP Violation in
Supersymmetric Model with Non-degenerate $A$-terms}

\maketitle

\begin{center}
\textsc{Shaaban Khalil} \\ \vspace*{1mm} \small{\textit{Ain Shams University,
Faculty of Science, Cairo 11566, Egypt}} \\
\vspace*{5mm} \textsc{Tatsuo Kobayashi} \\
\vspace*{1mm} \textit{Department of Physics, High Energy Physics
Division\\
    University of Helsinki \\
and\\ Helsinki Institute of Physics, P.O. Box 9 (Siltavuorenpenger
20 C)\\
    FIN-00014 Helsinki, Finland }\\
\vspace*{5mm} \textsc{Antonio Masiero} \\ \vspace*{1mm}
\textit{International School for Advanced Studies, SISSA,\\ via
Beirut 2-4,I-34100 Trieste, Italy.}\\
\end{center}

\vspace*{5mm}

\begin{abstract}
We study the CP phases of  the soft supersymmetry breaking terms in
string-inspired models with non-universal trilinear couplings. We
show that such non-universality plays an important role on all CP
violating processes. In particular these new supersymmetric sources of CP
violation may significantly contribute  to the observed CP phenomena in kaon
physics while respecting the severe bound on the electric dipole moment of
the neutron.
\end{abstract}

\renewcommand{\baselinestretch}{1.2} \large\normalsize

\section{Introduction}
CP phenomenology is sensitive to new physics beyond the standard model.
In supersymmeteric models, there appear new CP violating phases which
arise
from the complexity of the soft supersymmetry (SUSY) breaking terms,
i.e. the trilinear scalar $A$-terms ,  the bilinear scalar $B$-term and
the Majorana gaugino masses, as well as from the $\mu$ parameter.
The presence of these phases would give large contributions, e.g.,
to the electric dipole moment of the neutron (EDMN)
\cite{edmn1,edmn2,dugan,edmn3,edmn4,edmn5}
and to the CP violation parameters ($\varepsilon$ and
$\varepsilon'$ ) of the $K-\bar{K}$ system \cite{epsk,masiero}.
There has been a considerable
amount of work concerning these phases in the minimal supersymmetric
standard model (MSSM) \footnote{For a review, see e.g. \cite{cprev}
and references therein.}.
It was shown that to suppress the EDMN, either
large scalar masses (approaching more than 1 TeV) or
small CP phases (of order $10^{-3}$,
when all SUSY masses are of order 100 GeV) are required.
In the latter case,
 the MSSM SUSY phases generate
CP violation in the $K-\bar{K}$ system far below the experimental value.
Thus, the Cabibbo-Kobayashi-Maskawa (CKM) phase must provide
almost the whole contrbituion to the  observed CP violation in the
$K$-system.
\vskip 0.3cm
Recently the question whether the EDMN actually forces the SUSY
contributions to CP violation in the $K$ system to be quite small has been
vigorously readdressed. In particular this has been due to a change
in the perspectives of the SUSY model-building. While in the 80's and
early 90's most emphasis was put on the minimal SUSY extension of
the SM (i.e. the MSSM), more recently it has become clear that the
MSSM represents a very particular choice of SUSY extension of the SM
with drastic assumptions on the SUSY breaking terms. The advent of
superstring inspired model has even more stressed the particular
nature of the MSSM and the difficulty, in general, to obtain all the
strict boundary conditions on which the MSSM relies. If one gives up
the MSSM and goes for more general SUSY realizations it is possible to
avoid the above obstruction on large SUSY contributions to $\varepsilon$.
\vskip 0.3cm
Three ways out have been indiduated so far. First, even remaining within
the MSSM context, the complete computation of the SUSY contributions to the
EDMN involves several contributions and possible destructive interferences
can occur in some regions of the SUSY parameter space \cite{edmn4, edmn5}.
A second possibility occurs in the so-called models of effective
supersymmetries where the sfermion of the first two generations are very
heavy (in the tens of TeV range) while those of the third generation
remain light. Here the SUSY contributions to the EDMN are suppressed even
with the maximal SUSY phases either because the squarks in the loop
are very heavy or because the mixing angles are very small \cite{giudice}.
Finally, we come to the way out which is of most immediate interest in our
work. It relies on the non-universality of the trilinear $A$ terms of the soft
breaking sector of the SUSY Lagrangian \cite{abel}. Let us expand more on
this latter possibility.
\vskip 0.3cm
   In most of analysis  universal or degenerate $A$-terms have
been assumed, i.e., $(A_{U,D,L})_{ij}=A$ or $(A_{U,D,L})_{ij}=A_{U,D,L}$.
This is certainly a nice simplifying assumption,
but it removes some interesting degrees of freedom.
For example, every $A$-term would, in general have an independent CP
phase,
and in principle we would have $27(=3 \times 3 \times 3)$ independent
CP phases.
However, in the universal assumption only one independent CP phase
is allowed.
\vskip 0.3cm
The situation drastically changes if we are to allow for non-degenerate
{A} terms with different and independent CP phases.
For example, the off-diagonal element of the squark (mass)$^2$ matrix,
say $(M_Q^2)_{12}$, includes the term proportinal to
$(A_U)_{1i}(A_U^{\dag})_{i2}$.
However, in the universal or the degenerate case this term is always
real.
Furthermore, these off-diagonal elements play an important
role in $\varepsilon_K$, as shall be shown later.
If these  terms enlarge the  imaginary part of $(M_Q^2)_{12}$, CP violation
in the $K$-system may be enhanced. That such a case may occur was recently
shown  in Ref.\cite{abel}.
\vskip 0.3cm
In this paper we study more explicitly and concretely such aspects of the
CP violation in the $K$-system due to nondegenerate $A$-terms,
using soft SUSY breaking parameters derived
from superstring models with certain assumptions \cite{ST-soft,BIM}.
We assume real Yukawa matrices in order to study the CP violation effect
due to SUSY CP phases \footnote{In higher dimensional field theory and
some type of string compactification CP is a nice symmetry,
that is, Yukawa couplngs are real \cite{lim}.}.
Also we use generically realistic Yukawa
matrices. We investigate how much the CP violation parameter
$\varepsilon$ in the $K$-system is enhanced by the effect due to
nondegenerate $A$-terms.
\vskip 0.3cm
    The paper is organized as follows. In section 2 we review the
soft SUSY breaking terms derived from superstring models. We assign
family-dependent modular weights in order to have nondegenerate trilinear
couplings.
In section 3, we study the effect of these phases in the CP violating
physics of kaons. We show that SUSY CP phases could contribute to the
observed value of $\varepsilon$ in the $K$-system.
Section 4 deals with the EDMN.  We give our
conclusions in section 5.

\section{Soft SUSY breaking terms}

 First we give a brief review on the soft SUSY
breaking terms in string models,
\begin{eqnarray}
-\mathcal{L}_{\rm SB} &=& \frac{1}{6} \,Y^A_{ijk}\,\phi_i \phi_j
\phi_k + \frac{1}{2} \,(\mu B)^{ij}\,\phi_i \phi_j + \frac{1}{2}
\,(m^2)^{j}_{i}\,\phi^{*\,i} \phi_j+ \frac{1}{2} \,M_a\,\lambda
\lambda+\mbox{H.c.}~
\end{eqnarray}
where $Y^A_{ijk}=(YA)_{ijk}$,
the $\phi_i$ are the
scalar parts of the chiral superfields $\Phi_i$ and $\lambda$ are the
gauginos.
We assume the string model to have
 the same massless matter content of the MSSM, \ie\ three families of
quark doublets $Q_i$, the up-type quark singlets $U_i$, the down-type
 quark singlets $D_i$, lepton doublets $L_i$ and lepton singlets $E_i$
as well as two Higgs fields, $H_1$ and $H_2$.
Here we consider orbifold models with the overall moduli
field $T$ as well as the dilaton field $S$. We assume that the dilaton and
the
moduli fields contribute to SUSY breaking and the vacuum energy vanishes.
\vskip 0.3cm
In this case the soft scalar masses $m_i$ and the gaugino masses $M_a$
are written as \cite{BIM}
\begin{eqnarray}
m^2_i &=& m^2_{3/2}(1 + n_i \cos^2\theta),
\label{scalar}\\
M_a &=& \sqrt{3} m_{3/2} \sin\theta e^{- i \alpha_{S}},
\label{gaugino}
\end{eqnarray}
where $m_{3/2}$ is the gravitino mass,
$n_i$ is the modular weight of the chiral multiplet and $\sin \theta$
corresponds to a ratio between $F$-terms of $S$ and $T$.
For example, the limit $\sin \theta \rightarrow 1$ corresponds to
the dilaton-dominant SUSY breaking.
Here the phase $\alpha_S$ is originated from the $F$-term of $S$.
In the equation for $M_a$ the $T$-dependent threshold corrections are
neglected.
These latter are important only for the case where the tree level value is
very small, i.e., $\sin \theta \rightarrow 0$.
Here we do not discuss such limit.
Similarly the  $A$-terms are  written as
\begin{eqnarray}
A_{ijk} &=& - \sqrt{3} m_{3/2} \sin\theta e^{-i \alpha_s}
- m_{3/2} \cos\theta
(3 + n_i + n_j + n_k) e^{-i \alpha_T},
\label{trilinear}
\end{eqnarray}
where $n_i$, $n_j$ and $n_k$ are modular weights of fields in the
corresponding Yukawa coupling $Y_{ijk}$.
Here the phase $\alpha_T$ is originated from the $F$-term of
$T$ \footnote{We treat $\alpha_S$ and $\alpha_T$ as free paremeters.
If we fix a form of the SUSY breaking superpotential, these magnitudes
can be fixed \cite{bailin}.}.
If $Y_{ijk}$ depends on $T$, there appears another contribution.
However, we do not take such case.
\vskip 0.3cm
Thus, the gaugino masses and the  $A$-terms as well as the $B$-term are,
in general, complex.
We have a degree of freedom to rotate $M_a$ and $A_{ijk}$ at
the same time~\cite{dugan}.
Here we use the basis where $M_a$ is real.
In $A$-terms of the above basis, there remains only one independent
degree of freedom of the phase, i.e.,
$\alpha' \equiv \alpha_T - \alpha_S$.
However, note that in general $A$-terms can have different phases each
other
except the case with $\cos \theta \sin \theta =0$.
The case with $\cos \theta =0$ corresponds
to the dilaton dominant SUSY breaking leading to the universal $A$-term,
while the case with $\sin \theta =0$ corresponds to the moduli-dominant
SUSY breaking, where CP phases are universal, i.e.,
$A_{ijk}=\vert A_{ijk}\vert e^{i\alpha'}$.
\vskip 0.3cm
In order to avoid any conflict with the experimental results on
flavor
changing
neutral current processes , we assume that the soft scalar masses of the first
and
second families are degenerate, that is, the first and second families
have the same modular weights.
Under this assumption, we in general have the $A$-parameter matrix,
\begin{eqnarray}
A^{u,d}_{ij} = \left (
\begin{array}{ccc}
a_{u,d} & a_{u,d} & b_{u,d}\\
a_{u,d} & a_{u,d} & b_{u,d} \\
b'_{u,d} & b'_{u,d} & c_{u,d}
\end{array}
\right),
\label{Atex}
\end{eqnarray}
that is, all of the entries in the first $2 \times 2$ block are
degenerate,
and the (1,3) and (2,3) ((3,1) and (3,2)) entries are degenerate each
other.
After  assigning specific modular weights , we obtain explicit
values for the entries in
 the $A$-parameter matrix.
\vskip 0.3cm
In addition to the  soft terms, we have to fix the Yukawa matrices to
be able to perform an explicit computation.
There are several types of Ans\"atze for realistic Yukawa matrices.
Some typical Yukawa matrices leading to approximate values of quark
masses
and their mixing angles are enough for our purpose.
Here we assume 1)  every entry in the Yukawa matrix is real,
2) the Yukawa matrix is symmetric and
3) the Yukawa matrix has the following hierarchical structure,
\begin{eqnarray}
Y_{33}>Y_{ij}, \quad Y_{22}>Y_{mn},
\label{assum3}
\end{eqnarray}
where $Y_{i,j}$ is any entry except $Y_{33}$ and $Y_{mn}$ denotes
the (1,1), (1,2) and (2,1) entries.
\vskip 0.3cm
Under these assumptions, we can write a generic form of the down-Yukawa
matrix,
\begin{eqnarray}
Y^d_{ij}=Y^d_{33}
\left (
\begin{array}{ccc}
(m_d/m_b)\Theta^d_{11} & V^{CKM}_{12}(m_s/m_b) &
V^{CKM}_{13}\Theta^d_{13} \\
V^{CKM}_{12}(m_s/m_b)  & (m_s/m_b)  &
V^{CKM}_{23}\Theta^d_{23} \\
V^{CKM}_{13}\Theta^d_{13} & V^{CKM}_{23}\Theta^d_{23} & 1
\end{array}
\right),
\label{ykwd}
\end{eqnarray}
and the up-Yukawa matrix,
\begin{eqnarray}
Y^u_{ij}=Y^u_{33}
\left (
\begin{array}{ccc}
(m_u/m_t)\Theta^u_{11} & \sqrt{(m_um_c/m_t^2)}\Theta^u_{12}\Theta^u_{22}
&
V^{CKM}_{13}\Theta^u_{13} \\
\sqrt{(m_um_c/m_t^2)}\Theta^u_{12} \Theta^u_{22}  &
(m_c/m_t)\Theta^u_{22}  &
V^{CKM}_{23}\Theta^u_{23} \\
V^{CKM}_{13}\Theta^u_{13} & V^{CKM}_{23}\Theta^u_{23} & 1
\end{array}
\right),
\label{ykwu}
\end{eqnarray}
in terms of the eight free parameters,
$\Theta^{u,d}_{11}$, $\Theta^{u,d}_{12}$, $\Theta^{u,d}_{13}$ and
$\Theta^{u,d}_{23}$, while $\Theta^u_{22}$ is of order one.
In addition we have a constraint, $\Theta^d_{23}-\Theta^u_{23} \approx 1$.
A  detailed discussion of
this parametrization will be given in Ref.\cite{STA2} .
\vskip 0.3cm
Actually, most of symmetric and hierarchical Yukawa mass matrices which
have been already proposed in the literature are included in the above
textures (7) and (8). For example, five types of symmetric Yukawa matrices with five
texture zeros have been obtained in Ref.\cite{RRR}.
The following parameter assignments correspond to four
Ramond-Roberts-Ross (RRR) types,
\begin{eqnarray}
(\Theta^d_{23},\Theta^d_{13},\Theta^u_{23},\Theta^u_{13},\Theta^u_{12})
&=& (1,0,0,0,1) {\rm \ in \ the \ first \ RRR \ type},\\
&=& (1,0,0,1,0) {\rm \ in \ the \ third \ RRR \ type},\\
&=& (0,0,1,0,1) {\rm \ in \ the \ fourth \ RRR \ type},\\
&=& (0,0,1,1,0) {\rm \ in \ the \ fifth \ RRR \ type},
\end{eqnarray}
with the other prameters $\Theta^{u,d}_{ij}$ suppressed \footnote{The
second
RRR type corresponds to the suppressed value of $\Theta^u_{22}$.}.
Moreover, in Ref.\cite{KZ} string-inspired realistic quark mass matrices
have been studied and the obtained matrices correspond to
the case with $\Theta^d_{23}=\Theta^u_{12}=1$, a small value of
$\Theta^u_{23}$ and the other  $\Theta^{u,d}_{ij}$ suppressed.\\

\section{CP violation}
As an example of non-universal cases, we take $n_i= -1$ for the third
family
and $n_i=-2$ for the first and second families.
Also we assume that
$n_{H_1}=-1$ and $n_{H_2}=-2$.
In this case, we find the following texture for the $A$-parameter matrix
at the string scale
\begin{equation}
A^{d} = \left (
\begin{array}{ccc}
a_d & a_d & b_d\\
a_d & a_d & b_d \\
b_d & b_d & c_d
\end{array}
\right), \quad
A^{u} = \left (
\begin{array}{ccc}
a_u & a_u & b_u\\
a_u & a_u & b_u \\
b_u & b_u & c_u
\end{array}
\right),
\end{equation}
where
\begin{eqnarray}
a_u&=& m_{3/2}(-\sqrt{3} \sin\theta + 3 e^{-i \alpha'} \cos\theta),\\
a_d&=&b_u= m_{3/2}(-\sqrt{3} \sin\theta + 2 e^{-i \alpha'} \cos\theta),\\
b_d&=&c_u= m_{3/2}(-\sqrt{3} \sin\theta + e^{-i \alpha'} \cos\theta),\\
c_d&=&-\sqrt{3}m_{3/2}\sin\theta.
\end{eqnarray}
In this paper we take
$\Theta^d_{23}=\Theta^d_{13}=\Theta^u_{13}=1$ and
$\Theta^d_{11}=\Theta^u_{11}=\Theta^u_{12}=\Theta^u_{23}=0$ as an example.
Having specified the values of the soft terms at the string scale,
we can use the electroweak breaking conditions at $M_Z$, which at tree
level can be expressed as,
\begin{equation}
\mu^2 = \frac{m^2_{H_1} - m^2_{H_2}\tan^2\beta}{\tan^2\beta - 1} -
M^2_Z/2 ,
\label{electro1}
\end{equation}
\begin{equation}
\sin 2\beta = \frac{-2 \vert B.\mu \vert }{m^2_{H_1} + m^2_{H_2} + 2
\mu^2} ,
\label{electro2}
\end{equation}
where $\tan \beta$ is the ratio of the vacuum expectation values of the Higgs
fields. Using eqs.(\ref{electro1}) and (\ref{electro2})
we can determine the value of $\vert\mu\vert$ and $\vert B \vert$ as
functions
on $m_{3/2}$, $\theta$ and $\alpha'$. We impose $\phi_B=0$ to avoid large
EDMs. The origin of this latter phase is linked to the way the $\mu$ term
is produced in effective supergravities. Given the focus of our paper on
the role of non-universal $A$-terms for CP violation, we are not going to
discuss the different mechanisms to originate $\mu$ and we simply consider
a vanishing $\phi_B$.
We also assume a low value of $\tan \beta$, namely we consider it to be
of order 3. Then
all the supersymmetric particle spectrum is completely determined in
terms of $m_{3/2}$,
$\theta$ and $\alpha'$.
Apart from the key-role of the non-universal A-terms  and of their
phases in CP violation, it is worth mentioning that
they have also important effects on
the SUSY spectrum and on the non-CP violating processes~\cite{edmn5}
and~\cite{falk}
\vskip 0.3cm
Before computing the contribution to CP violation in the kaon system from
the phases in the above matrices $A^{d}$ and $A^{u}$, one important remark
is in order. The fact of having non-degenerate A-terms is essential to
obtain non-vanishing CP violating contributions from SUSY loops when the
CKM matrix is taken to be real. Namely, it can be shown that, independently
from how large one takes the SUSY CP phases $\phi_A$ and $\phi_B$, in the
presence of degenerate A-terms and real CKM there is no way of generating
a phenomenologically viable amount of CP violation in K physics. This
point was already emphasized by Abel and Frere \cite{abel} who showed
explicitly that $\varepsilon$ turns out to be extremely tiny in box-diagrams
with chargino/up-squark exchange and external left-handed quarks when the
matrices $(YA)$ in our notation are symmetric ( notice that the symmetric
form of the trilinear matrcies $A^u$ and $A^d$ and the Yukawas matrices $Y^u$ and
$Y^d$ in our case is at the GUT scale and indeed at weak scale our trilinear
scalar terms are not symmetric in generation space since they have different
running). Their proof can be readily extended to SUSY loops with gluino exchange .
This can be seen as follow.
\vskip 0.3cm
 The value of the indirect CP violation in the Kaon
decays, $\varepsilon$, is defined as
\begin{equation}
\varepsilon = \frac{e^{i\frac{\pi}{4}}
{\rm Im} M_{12}}{\sqrt{2} \Delta m_K},
\end{equation}
where $\Delta m_K= 2 {\rm Re} \langle K^0 \vert H_{eff} \vert \bar{K}^0
\rangle = 3.52 \times 10^{-15}$ GeV.
The amplitude
$M_{12}=\langle K^0 \vert H_{eff} \vert \bar{K}^0 \rangle$ is given
in Ref.\cite{masiero} in terms of the mass insertion $\delta_{AB}$
defined
by $\delta_{AB} = \frac{\Delta_{AB}}{\tilde{m}^2}$ where $\tilde{m}$ is
an
average sfermion mass and the $\Delta$'s denote off-diagonal, flavor
changing terms in the sfermion mass matrices.
\vskip 0.3cm
We consider gluino exchange contributions with all external left-handed
quarks. In this case the relevant flavor changing mass insertions that
appear on the internal squark propagator lines
accomplish the transition from $\tilde{d}_{1L}$
to $\tilde{d}_{2L}$ ($1$ and $2$ are flavor indices):
\begin{equation}
(\Delta_{LL}^d)_{12} = \left[ K M_{\tilde{Q}}^2 K^{\dag} \right]_{12},
\end{equation}
where $K$ is the Kobayashi-Maskawa matrix.
\vskip 0.3cm
In the case of degenerate A-terms, i.e. $A_{ij}=A \delta_{ij}$,
one obtains\cite{epsk}
\begin{equation}
(\Delta_{LL}^d)_{12} \simeq -\frac{1}{8\pi^2}
(K^{\dag} h_U^2 K )_{12} ( 3m_0^2 + \vert A \vert^2),
\end{equation}
\ie, the flavor changing mass insertions remain real (for real CKM matrix)
independently from the phase $\phi_A$.
In such diagonal or degenerate cases the SUSY
contribution to  CP violation relies on  the $(\delta_{12})_{LR}$
and $(\delta_{12})_{RL}$ mass insertions and it turns out to be very
small \cite{abel}. Obviously this is no longer true if we switch on the
CP violating phase of the CKM matrix. In that case SUSY loops can give a
non-negligible contribution to $\varepsilon_K$, although such contribution
cannot be the major source of CP violation.
\vskip 0.3cm
To obtain a  large  SUSY contribution to $\varepsilon$, it is necessary to
enhance the values of  ${\rm Im}(\delta_{12})_{LR}$ and ${\rm
Im}(\delta_{12})_{RL}$. The non-degenerate A-terms is an interesting
example for enhancing these quantities since the off-diagonal terms,
namely $A^d_{12}$, lead
to non vanishing value of $(\delta_{12})_{LR}$ and $(\delta_{12})_{RL}$
at the tree level (see eq.(1)).
\vskip 0.3cm
We consider the box diagrams which are responsible for the
$K^0-\bar{K}^0$ transition and we focus on the contribution coming from
gluino exchange in the loop (as we shall see below this turns out to be
the dominant contribution to $\varepsilon$) \cite{masiero}:
\begin{eqnarray}
        M^{\rm gluino}_{12}&=&-\frac{\alpha_S^2}{216
m_{\tilde{q}}^2}\frac{1}{3}
m_{K}f_{K}^{2}
        \Biggl\{  \left(\delta^d_{12}\right)^2_{LL}
        \left(  24\,x\,f_6(x) + 66\,\tilde{f}_6(x) \right)
        \nonumber \\
        &+&  \left(\delta^d_{12}\right)^2_{RR}
        \left(  24\,x\,f_6(x) + 66\,\tilde{f}_6(x) \right)
        \nonumber \\
        &+&
\left(\delta^d_{12}\right)_{LL}\left(\delta^d_{12}\right)_{RR}
        \left[(384(\frac{m_{K}}{m_{s}+m_{d}})^{2} + 72)
        \,x\,f_6(x)  \right. \nonumber \\
        &&+ \left. (-24 (\frac{m_{K}}{m_{s}+m_{d}})^{2}
        + 36) \tilde{f}_6(x) \right]
        \nonumber \\
        &+&  \left(\delta^d_{12}\right)^2_{LR}
        \left[  -132 \left(\frac{m_{K}}{m_{s}+m_{d}}\right)^{2} x\,
f_6(x) \right]
        \nonumber \\
        &+&  \left(\delta^d_{12}\right)^2_{RL}
        \left[  -132 \left(\frac{m_{K}}{m_{s}+m_{d}}\right)^{2} x\,
f_6(x) \right]
        \nonumber \\
        &+&
\left(\delta^d_{12}\right)_{LR}\left(\delta^d_{12}\right)_{RL}
        \left[ -144 \left(\frac{m_{K}}{m_{s}+m_{d}}\right)^{2} -84
\right]
        \tilde{f}_6(x)
        \Biggr\}.
\end{eqnarray}
Here, $x=(\frac{m_g}{M_{\tilde{q}}})^2$ and the functions $f_6(x),
\tilde{f}_6(x)$ are given in Ref.\cite{masiero}.  The above result is
obtained in the so-called superKM basis \cite{hall} by making use of the
mass insertion approximation method.
\vskip 0.3cm
As we mentioned,
we  assume that the CKM matrix is real and the soft SUSY
breaking terms are the only source for the complexity of the
amplitude $M_{12}$. The relevant contribution to CP violation comes from
the terms proportional to $(\delta_{12})_{LR}$ and
$(\delta_{12})_{RL}$ in the above expression. Going to the basis where the
down quark mass matrix is diagonal, the   mass insertion
$(\delta_{12})_{LR}$ is given by:
 \begin{equation}
(\delta^d_{12})_{LR} = U_{1i} (Y^A_d)_{ij} U^T_{j2},
\end{equation}
where $U$ is the matrix  diagonalizing the symmetric down quark mass matrix.
The most relevant contributions in the above equation are
\begin{equation}
(\delta^d_{12})_{LR} \simeq U_{11} (Y^A_d)_{12} U^T_{22} + U_{12}
(Y^A_d)_{22}
U^T_{22} + U_{13}
(Y^A_d)_{33} U^T_{23},
\end{equation}
which implies that Im$(\delta^d_{12})_{LR}$ is of the same order as
Im$(Y^A_d)_{12}$
and indeed it is found to be of order $10^{-4}$ and the same for
$(\delta_{12})_{RL}$. Moreover the values of Im $(\delta_{12})_{LL}$ and
Im$(\delta_{12})_{RR}$ are non zero unlike in the universal case,
but are smaller than
$(\delta_{12})_{LR}$.
\vskip 0.3cm
Also we estimate the chargino contribution to
$\varepsilon$. It is found that it
is proportional to $(\delta^u_{13})_{LR} (\delta^u_{23})_{LR}$. The
amplitude
for the chargino box contribution to $K^0-\bar{K}^0$ mixing can be
written as
\begin{equation}
M^{\rm chargino}_{12}\simeq -\frac{K_{13} K_{23} \alpha_W}{216
m_{\tilde{q}}^2}\frac{1}{3} m_{K}f_{K}^{2} (\delta^u_{13})_{LR}
(\delta^u_{23})_{LR} \left [-132
\left(\frac{m_{K}}{m_{s}+m_{d}}\right)^{2} x\, f_6(x)\right],
\end{equation}
where $x=(\frac{m_{\chi^{\pm}}}{M_{\tilde{q}}})^2$. The values of
$(\delta^u_{13})_{LR} (\delta^u_{23})_{LR}$ are two order of
magnitude larger than the values of $(\delta_{12})^2_{LR}$ but because of
the smallness
of the coupling it is found that the amplitude of the chargino
contribution is one
order of magnitude less than the gluino amplitude. Thus the main
contribution to
$\varepsilon$ is due to the gluino exchange. However, this is not
necessarily true in
the case of non-vanishing $\phi_B$.
\vskip 0.3cm
 Using the above values of the mass
insertions we can
determine the SUSY contribution to  $\varepsilon$. Fig. 1 shows
$\varepsilon$ in
terms of $\sin \alpha'$ for $\theta \simeq 0.8$ rad. and $m_{3/2} \simeq
100$ GeV.
\vskip 0.3cm
\begin{figure}[h]
\psfig{figure=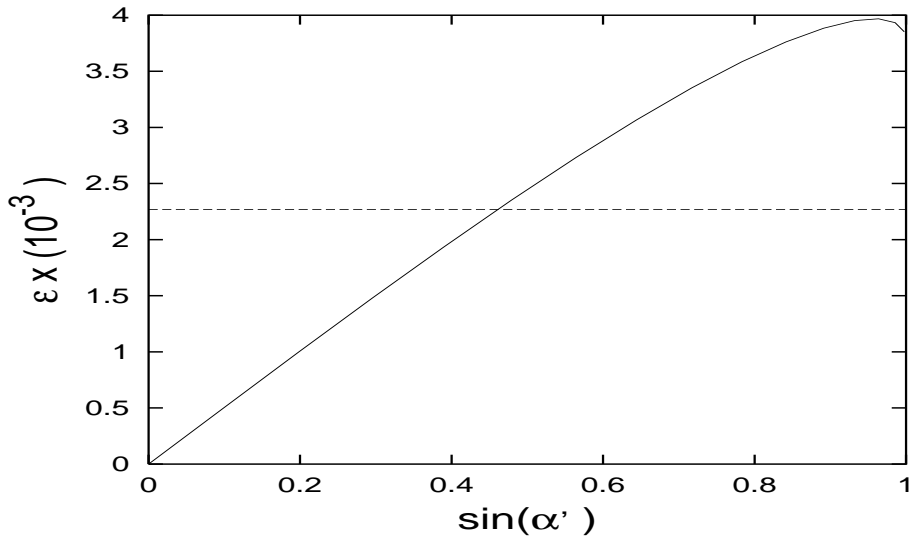,height=7cm,width=12cm}
\caption{The values of $\varepsilon $ versus $\sin \alpha'$ where
the goldstino angle $\theta \simeq 0.8$ rad. and $m_{3/2} \simeq 100$
GeV.}
\end{figure}
\vskip 0.3cm
Also we give the values of $\varepsilon $ in terms of $\theta$ in
Fig. 2, which shows that the non-universality between the soft
supersymmetry breaking terms is preferred to enhance the SUSY CP violating
contribution.
\vskip 0.3cm
\begin{figure}[h]
\psfig{figure=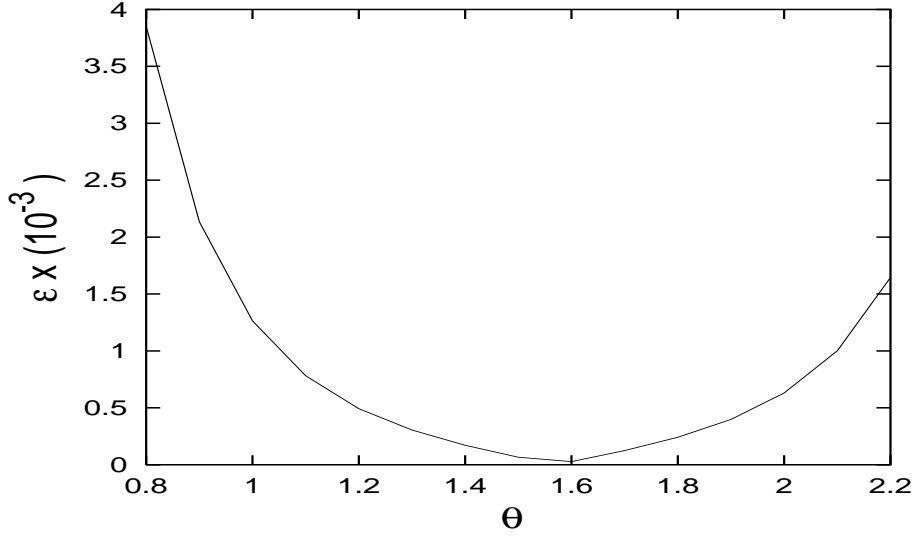,height=7cm,width=12cm}
\caption{The values of $\varepsilon $ versus $\theta$ where
$\alpha' \simeq \pi/2$ and $m_{3/2} \simeq 100$
GeV.}
\end{figure}
\vskip 0.3cm
Finally we present the values of $\varepsilon $ as a function of the
gravitino mass in Fig. 3. \\
\vskip 0.3cm
\begin{figure}[h]
\psfig{figure=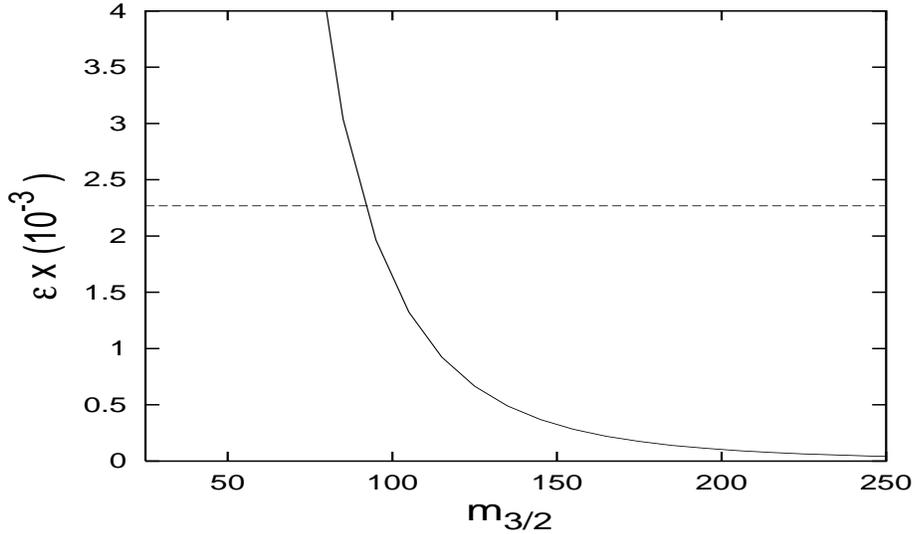,height=7cm,width=12cm}
\caption{The values of $\varepsilon $ versus $m_{3/2}$ where
the goldstino angle $\theta \simeq 0.8$ rad. and $\alpha'\simeq \pi/2$}
\end{figure}
\vskip 0.3cm
It is interesting to note that for $m_{3/2}\sim  100$ GeV we
obtain large values of $\varepsilon $, which even exceed the
experimental limit $2.2\times 10^{-3}$. Thus we have a constraint
on  $(m_{3/2}, \alpha')$ from the experimental limit on
$\varepsilon $. For instance, in case of $\alpha'=\pi/2$ we find
that $m_{3/2}> 120$ GeV.
\vskip 0.3cm
We can proceed analogously
for other values of $\Theta^{u,d}_{ij}$. For example, the
Ramond-Roberts-Ross textures  lead to very similar results.
 As another exmaple, we can take the case with
$n_{H1}=-1$, $n_{H2}=-2$ and $n_i=-1$ for the other matter fields.
This case leads to a degenerate $A$-matrix, i.e. $A^u_{ij}=A^u\delta_{ij}$
and $A^d_{ij}=A^d\delta_{ij}$. However, note that these $A$-parameters in
general have CP phases independent of the gaugino mass unlike the
case where every field has the same modular weight $n_i=-1$. Consistently
with our previous general considerations on the necessity of non-degeracy
of the A matrices to have sizable CP violating contributions, in this case
$\varepsilon$ turns out to be smaller than $O(10^{-3})$.
\section{Electric dipole moment of the neutron}
The supersymmetric contributions to the  EDMN include gluino,
chargino and  neutralino loops..
Since we are considering the case with vanishing
$\phi_B$, the gluino contribution is dominant. For the   EDM of the
quark $u$ and $d$ it amounts to~\cite{masiero}
\begin{equation}
d^g_d/e =
-\frac{2}{9}\frac{\alpha_S}{\pi}\frac{m_{\tilde{q}}}{M_{\tilde{q}}^2}\
M_1(x)\ {\rm Im}(\delta^d_{11})_{LR},
\end{equation}
\begin{equation}
d^g_u/e = \frac{4}{9}\frac{\alpha_S}{\pi}
\frac{m_{\tilde{q}}}{M_{\tilde{q}}^2}\
M_1(x)\ {\rm Im}(\delta^u_{11})_{LR},
\end{equation}
where $m_{\tilde{g}}$ is the gluino mass, $M_{\tilde{q}}^2$ is the
average squark mass. The function $M_1(x)$ is given by
\begin{equation}
M_1(x)= \frac{1+4 x -5 x^2 +4 x \ln(x) + 2 x^2 \ln(x)}{2(1-x)^4}.
\end{equation}
As we explained in the last section, by using the electroweak breaking
condition we can write all the spectrum
in terms of $m_{3/2}$, $\theta$, $\alpha_S$, and  $\alpha_T$.
Then the EDM $d^g_q/e$ is given in terms of these parameters.
We use  the non-relativistic quark model approximation of
the EDMN :
\begin{equation}
d_n = \frac{1}{3}(4 d_d - d_u).
\end{equation}

The mass insertion $(\delta_{11})_{LR}$ is given by
$$(\delta_{11}^d)_{LR}= (U^T Y^A_d U)_{11} \simeq U_{21}
(Y_{21}A_{21}^d) U_{11} + U_{21} (Y_{22}A_{22}^d) U_{21}.$$
In our case,  we find that the $\delta_{11}$ is only one order of
magnitude less than the $\delta_{12}$. Unless the phases appearing in
$(\delta_{11}^d)_{LR}$ are small, we would expect the imaginary part of
this latter quantity
to be of
order $10^{-5} - 10^{-6}$. This implies that the gluino
contribution to the EDMN in this model is of order of
$10^{-25}e~cm$ . Recently it has been shown that the above EDMN
contributions can interfere destructively with other contributions in some
regions of the SUSY
parameter space \cite{edmn4,edmn5}. In section 3, we have shown that there
exists a relatively large region of the
 the parameter space to enhance
$\varepsilon$  with non-universal $A$-parameters. Thus, it may be possible
to find some cancellation with other contributions allowing for the EDMN
to be suppressed to values below
 $10^{-25}e cm$. To find such a parameter
space, a detailed analysis is needed  in particular with the  inclusion of
the
 effects of $\phi_B$ (we plan to provide it elsewhere~\cite{STA2}). Other than such
cancellation, as we said, the non-universal cases
could
 include the very specially fine-tuned case where only CP phases of
$A$-elements contributing to the EDMN are suppressed.
\vskip 0.3cm
Finally we  comment on the $\Delta S=1$
 CP violating parameter
$\varepsilon'/ \varepsilon$.
 In our case ,where
$(\delta_{12})_{LR}$ and $(\delta_{12})_{RL}$ give the important
contributions to the CP violation processes in kaon physics,  the relevant
part of  the effective hamiltonian $\mathcal{H}_{\rm^M eff}$
 for $\Delta S=1$  CP violation is
 \begin{equation}
\mathcal{H}_{\rm eff} = C_8 O_8 + \tilde{C}_8 \tilde{O}_8,
\end{equation}
where $C_8$ and $O_8$ are given in Ref.\cite{masiero} and
$\tilde{C}_8$ can be obtained from $C_8$ by exchange
$L\leftrightarrow R$ and the matrix element of the operator
$\tilde{O}_8$ is obtained from the matrix element of $O_8$
multiplying them by $(-1)$. Since we have $(\delta_{12})_{LR}$
approximately equal to $(\delta_{12})_{RL}$, then $C_8$ is
very close to $\tilde{C}_8$  and hence, the value of
$\varepsilon'$ is very small. This cancellation between the
different contributions to $\varepsilon'$ is mainly due to the
symmetric nature of the trilinear and/or Yukawa matrices we adopted.
It has recently been shown that in the absence of such cancellation it is
possible to obtain large values of $\varepsilon'/varepsilon$ (compatible with the
experimental results of NA31 and KTeV) using the flavour changing
trilinear scalar terms of the soft breaking
sector~\cite{murayama}.

\section{Conclusions}
We have studied CP violation in  the SUSY model with the
non-degenerate A-terms derived from superstring theory. This type
of non-universality has a significant effect in the CP violation.
We have shown the region of the parameter space where we have SUSY
contribution for $\varepsilon$ of order $10^{-3}$.
\vskip 0.3cm
It is interesting to investigate effects of non-universality on
other CP aspects, e.g. detailed analysis of the EDMN. Also the
effect of the non-degearcy of the A-terms is important in studying
the B-physics.
\vskip 0.3cm
We have considered the case where the dilaton field and only the
overall moduli field contribute to SUSY breaking and in this case
only one independent CP phase $\alpha'=\alpha_T-\alpha_S$ appears
in the $A$-parameters. It would be interesting to discuss
multi-moduli cases \cite{multi}, where several independent CP
phases appear.
\vskip0.75truecm
\begin{center}
\noindent{\small ACKNOWLEDGEMENTS}
\end{center}
\vskip0.5truecm
The authros would like to thank S. Bertolini and D.~Bailin for useful
discussions. This work was partially supported by the Academy of Finland
under the Project no. 44129.

\end{document}